\documentclass[msmath,amssymb,%
floatfix,superscriptaddress,a4paper]{article}

\usepackage{graphicx}
\usepackage{dcolumn}
\usepackage{bm}
\usepackage[latin1]{inputenc}
\usepackage{amssymb,amsmath,mathrsfs}
\usepackage{color,ulem}
\usepackage[margin=1.7cm]{geometry}
\usepackage[T1]{fontenc}
\usepackage{authblk}
\usepackage{cite}
\usepackage{array}
\usepackage{multirow}
\newcolumntype{L}[1]{>{\raggedright\let\newline\\\arraybackslash\hspace{0pt}}m{#1}}
\newcolumntype{C}[1]{>{\centering\let\newline\\\arraybackslash\hspace{0pt}}m{#1}}
\newcolumntype{R}[1]{>{\raggedleft\let\newline\\\arraybackslash\hspace{0pt}}m{#1}}

\begin{document}

\newcommand{\ittext}[1]{\mbox{\rm\scriptsize #1}}

\title{\bfseries Scale-invariant nonlinear optics in gases}

\author[1]{C.~M.~Heyl}
\author[1]{H.~Coudert-Alteirac}
\author[1]{M.~Miranda}
\author[1] {M.~Louisy}
\author[2,3]{K.~Kovacs}
\author[2,3]{V.~Tosa}
\author[3,4]{E.~Balogh}
\author[3,4]{K.~Varj\'u}
\author[1]{A.~L'Huillier}
\author[5]{A.~Couairon}
\author[1]{C.~L.~Arnold} 
\affil[1]{Department of Physics, Lund University, P. O. Box 118, SE-221 00 Lund, Sweden}
\affil[2]{National Institute for R\&D Isotopic and Molecular Technologies, Cluj-Napoca, Romania}
\affil[3]{ELI-ALPS, ELI-Hu Nkft, Dugonics ter 13, Szeged 6720, Hungary}
\affil[4]{Department of Optics and Quantum Electronics, University of Szeged, Dom ter 9, 6720 Szeged, Hungary}
\affil[5]{Centre de Physique Th\'eorique, \'Ecole Polytechnique, CNRS, F-91128, Palaiseau, France}
\renewcommand\Authands{ and }
\date{\vspace{-5ex}}

    \maketitle
{\bfseries
Nonlinear optical methods are becoming ubiquitous in many areas of modern photonics. They are, however, often limited to a certain range of input parameters, such as pulse energy and average power, since restrictions arise from, for example, parasitic nonlinear effects, damage problems and geometrical considerations.  
Here, we show that many nonlinear optics phenomena in gaseous media are scale-invariant if spatial coordinates, gas density and laser pulse energy are scaled appropriately. 
We develop a general scaling model for (3+1)-dimensional wave equations, demonstrating the invariant scaling of nonlinear pulse propagation in gases. 
Our model is numerically applied to high-order harmonic generation and filamentation as well as experimentally verified using the example of pulse post-compression via filamentation. Our results provide a simple recipe for up-or downscaling of nonlinear processes in gases with numerous applications in many areas of science.  
\vspace{0.5cm}
}   

Nonlinear interactions of intense short laser pulses with gaseous media form the basis behind a wealth of interesting phenomena such as multiphoton ionization \cite{MainfrayRPP1991} and plasma formation \cite{ConradsPSST2000}, spectral broadening (which can be used for pulse compression \cite{NisoliOL1997,HauriAPB2004,CouaironJModOpt2006}), harmonic generation and wave-mixing \cite{ReintjesPRL1976}, as well as the creation of attosecond pulses \cite{KrauszRMP2009} and the formation of electron or ion beams \cite{EsareyRMP2009}. 
Advances in femtosecond laser technology constantly yield shorter pulses, higher pulse energies, and higher repetition rates \cite{Danson2015,LimpertIEEE2014,FattahiOPA2014}.
However, to fully explore this newly available parameter regime, which gives access to e.g. faster time scales and higher intensities, is often challenging because of damage problems, additional (unwanted) nonlinear effects, or geometrical restrictions.  
We illustrate this challenge for two important applications of nonlinear optics, filamentation in gases used e.g. for laser pulse compression, and high-order harmonic generation (HHG) providing the basis for attosecond science.

The propagation of an intense short laser pulse in a transparent medium induces nonlinear effects caused e.g. by the intensity dependence of the refractive index. When self-focusing due to the Kerr effect balances defocussing caused by diffraction and plasma generation, a filament can be created.  
In addition, self-phase modulation and self compression may take place in the filament, resulting, possibly after further compression, in ultrashort pulses close to the fundamental limit of a single cycle \cite{CouaironOL2005}. Forming a filament requires a certain power, known as the critical power for self-focusing \cite{Marburger75,FibichOL2000}. At slightly higher power, limitations arise and multiple filaments are created \cite{MlejnekPRL1999}. 
Different attempts were suggested to increase the output energy \cite{VarelaOL2010,VarelaOE2009,suda05,CouaironOL2005,Fourcade_Dutin_OptLett_10,ArnoldNJP2010}. However, pulse compression using filaments (or similarly hollow fibers) is still limited to pulse energies of typically a few mJ \cite{BohmanOE08,SkupinPRE06}, which is approximately two to three orders of magnitude below the maximum pulse energies available from today's femtosecond laser sources. To scale up pulse post-compression into the 100~\,mJ-regime and above will enable the production of few-cycle pulses with unprecedented peak power, opening the door to new applications. 

Our second example pertains to high-order harmonic generation, which occurs when intense short laser pulses interact with a gas of atoms or molecules at an intensity of $\sim 10^{14}$\,W/cm$^2$ \cite{FerrayJPB1988}. This process leads to the formation of attosecond light pulses, which can be used for pump-probe studies of ultrafast electron dynamics \cite{KrauszRMP2009}.
A major limitation of attosecond science is the low photon flux available \cite{LeoneNatPho2014}. Since the early days, a strong effort has been devoted to optimize and scale up HHG \cite{LhuillierJPB1991, TakahashiPRA2002, PopmintchevSci2012, RudawskiRSI2013} aiming for an efficient conversion of high laser pulse energies into the extreme ultraviolet (XUV). 
In spite of this effort, propagation effects and geometrical considerations have limited the useful input laser pulse energy and only a few groups have employed pulse energies exceeding 10~mJ \cite{TakahashiOL2002,HergottPRA2002,RudawskiRSI2013,CassouOL2014,TzallasNatPhy2011}. 
In the opposite direction, progress in laser technology now enables the generation of laser pulses with  $\mu$J energies at MHz repetition rates \cite{ChiangAPL2012}. 
In this regime, macroscopic phase matching issues have limited the conversion efficiency into the XUV and only recent attempts point towards a solution of this problem \cite{HeylJPB2012,RothhardtNJP2014}. 
An efficient down-scaling of HHG will enable compact attosecond sources at high repetition rate, of great interest for studies of time-resolved photoelectron emission processes on surfaces and for coincidence spectroscopy in atoms and molecules.

Taken together, these examples illustrate the strong need for a general methodology that enables up- or downscaling of nonlinear processes in gases. 
Here we present such a methodology and introduce a set of general scaling relations directly derived from basic propagation equations for ultrashort laser pulses.
We verified our theoretical predictions by performing numerical propagation calculations for the two phenomena described above, filamentation and HHG. 
We show how these processes can be invariantly scaled to laser pulse energies well above the 100~mJ level, with no fundamental upper limit.
Moreover, we experimentally verified the invariant scalability of pulse compression via filamentation within a driving laser pulse energy range exceeding one order of magnitude.
Our scaling formalism is simple and general, and opens up completely new parameter regimes for nonlinear optics in gaseous media and more generally for ultrafast science. 

\subsection*{Scaling principles}
We illustrate our scale-invariant nonlinear optics framework using general wave equations.
Nonlinear pulse propagation in gases (including generation of new frequencies) is usually treated using wave equations in scalar and paraxial approximation, which can be directly derived from Maxwell's equations. Such wave equations describe electromagnetic waves propagating in one direction, exhibiting only small angles relative to the optical axis.
Without any limitation of the spectral bandwidth and thus of the minimum pulse duration, the propagation equation for the electrical field in frequency representation $\hat{E}(r,z,\omega) = \int_{-\infty}^{\infty} \exp(i\omega t) E(r,z,t) \mathrm{d}t$, usually referred to as the Forward Maxwell Equation \cite{HusakouPRL2001}, can be written as:
\begin{equation}
\bigg[\frac{\partial}{\partial z} - \frac{i}{2k(\omega,\rho)}\Delta_{\bot} - ik(\omega,\rho)\bigg] \hat{E} = \frac{i\omega^2}{2 k(\omega,\rho)c^2 \epsilon_0} \hat{P}_{\text{NL}}.
\label{eq:fme}
\end{equation}
Here, $k(\omega,\rho) = n(\omega,\rho)\omega/c$ denotes the wave number with angular frequency $\omega$, refractive index $n=n(\omega,\rho)$, and speed of light in vacuum $c$.
$\rho$ is the gas density, $\hat{P}_{\text{NL}}$ is the frequency representation of the nonlinear polarization induced by the electric field $E$ and $\epsilon_0$ is the vacuum permittivity.
For short pulse propagation, exact knowledge of the refractive index $n$, e.g. in the form of a Sellmeier equation is required.
For pulse propagation in the visible and near infrared spectral region, $k(\omega,\rho)$ is real-valued, but linear absorption can easily be included by a complex wave number.
For the sake of simplicity, we consider linear polarization and rotational symmetry and thus a single radial coordinate $r$, although our theory does not require these simplifications.
The transverse Laplace-operator in equation~\eqref{eq:fme} then becomes $\Delta_{\bot} = \partial^2/\partial r^2 + 1/r \cdot \partial/\partial r$.
Via the nonlinear polarization a large number of nonlinear interactions can be considered, such as self-focusing, self-phase modulation, field ionization, harmonic generation, and plasma-defocussing. 

\begin{figure}[t]
	\begin{center}
		\includegraphics[width=0.48\textwidth]{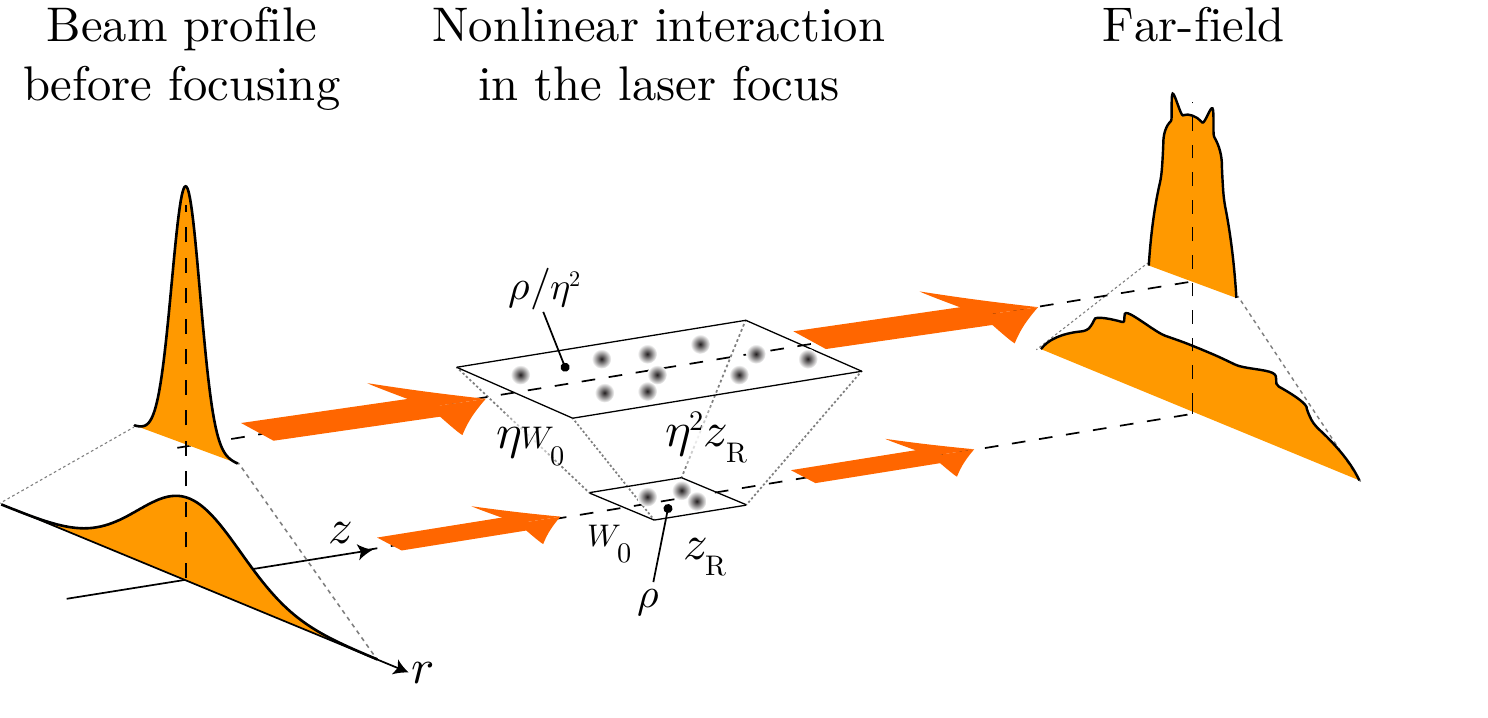}
		\setlength{\unitlength}{1mm}
	\end{center}
	\caption{\textbf{Scaling principle:} Illustration of scale-invariant nonlinear optics. A laser pulse is focused into a gas medium with length $L$ and density $\rho$. Nonlinear propagation effects lead to a modification of the spatiotemporal pulse profile (bottom). Identical spatiotemporal modifications can be expected if a more intense laser pulse is focused weaker (to reach the same intensity) into a larger medium with length $\eta^2 L$ and lower density $\rho/\eta^2$ (top). Note that in the up-scaled configuration, the far-field beam diameter is smaller and the amplitude larger at identical distance $z$ from the focus.}
	\label{fig:illustr}
\end{figure}

For propagation in vacuum, the right hand side of equation~\eqref{eq:fme} vanishes and $k(\omega,\rho) \rightarrow k(\omega,0)=\omega/c$.
We now introduce the field $\hat{\cal E} \equiv \hat{E}\exp[-i\omega z/c]$ and rewrite equation~\eqref{eq:fme}: 
\begin{equation}
\left[\frac{\partial}{\partial z} - \frac{ic}{2\omega}\Delta_{\bot}\right] \hat{\cal E} = 0.
\label{eq:g}
\end{equation}
The change of fields from $\hat{E}$ to $\hat{\cal E}$ formally corresponds to a transformation of equation~\eqref{eq:fme} from the laboratory frame to a frame moving at the vacuum speed of light $c$ \cite{GeisslerPRL99}.
It should be noted that $\hat{\cal{E}}$ is an electric field, not an envelope.
No envelope approximations and thus no restrictions on the spectral bandwidth are made.
Equation~\eqref{eq:g} is invariant under the following transformations:
$r \longrightarrow  \eta r$ and
$z \longrightarrow  \eta^2 z$ (see Table 1), 
where $\eta$ is a scaling parameter. If $\hat{\cal E}(r,z)$ is a solution to the wave equation, $\hat{\cal E}(r/\eta,z/\eta^2)$ is a solution as well.
For monochromatic waves, one prominent solution of equation~\eqref{eq:g} is the Gaussian beam. The scaling is obvious for the characteristic spatial parameters of the Gaussian beam, i.e. the beam radius $W_0$ and the Rayleigh length $z_{\mathrm R}$: $W_0 \longrightarrow \eta W_0$ and $z_{\mathrm R} \longrightarrow \eta^2 z_{\mathrm R}$.
While the Gaussian beam is just one possible solution to equation~\eqref{eq:g}, more generally, any kind of beam that can be described by this wave equation is scale-invariant under the above specified transformation.

These basic scaling principles can be generalized to ultra-short laser pulse propagation in gases and a wide range of nonlinear interactions, if the medium density and the input laser pulse energy $\varepsilon_{\textrm {in}}$ are included as scaling parameters. 
By introducing $\hat{\cal{E}}$ and $\hat{\cal P}_{\text{NL}} \equiv  \hat{P}_{\text{NL}} \exp[-i\omega z/c]$ into equation~(\ref{eq:fme}), we obtain:
\begin{equation}
\bigg[\frac{\partial}{\partial z} - \frac{i}{2k(\omega,\rho)}\Delta_{\bot} - iK(\omega,\rho)\bigg] \hat{\cal E} = \frac{i\omega^2}{2 k(\omega,\rho)c^2 \epsilon_0} \hat{\cal P}_{\text{NL}}(\rho),
\label{eq:fme1}
\end{equation}
where $K(\omega,\rho) = k(\omega,\rho)-k(\omega,0)$ is proportional to $\rho$ and describes pulse dispersion [see supplementary information (SI) for details]. By neglecting the weak pressure dependence of $k(\omega,\rho)$ in the denominator of the diffraction term, the left hand side of equation~\eqref{eq:fme1} is invariant under the above transverse and longitudinal scaling transformations, if simultaneously the gas density is scaled, i.e. $\rho  \longrightarrow  \rho/\eta ^2$. 
Similarly, the nonlinear polarization and consequently the right hand side of equation~\eqref{eq:fme1} is proportional to gas pressure $p$ for a wide range of nonlinear interactions (throughout the manuscript, we assume $p \propto \rho$, taking into account a constant temperature). 
Finally, the input energy $\varepsilon_{\textrm{in}}$, proportional to the radial (and temporal) integral of the absolute square of the input field, needs to be scaled as $\varepsilon_{\textrm{in}}  \longrightarrow  \eta^2 \varepsilon_{\textrm{in}}$, to ensure that the 
field amplitude, which affects $\hat{\cal P}_{\text{NL}}$ is kept constant under the scaling transformation.
The output pulse energy $\varepsilon_{\textrm{out}}$, proportional to the integral of the absolute square of the field at the end of the medium, follows the same scaling: 
$\varepsilon_{\textrm{out}}  \longrightarrow \eta^2 \varepsilon_{\textrm{out}}$. This scaling applies as well to the generation of new frequencies, as shown for the case of HHG below.  
In practice, the geometrical scaling can be achieved by changing the focusing geometry (focal length and/or beam diameter before focusing) as well as the medium length. 
It should be noted that the transformation to the moving frame, leading to equation~\eqref{eq:fme1}, was performed to illustrate the scaling principles, but does not constitute a general limitation of the formalism. The scaling itself is independent from the reference frame.

According to the above relations (see also Table 1), any spatiotemporal modifications of the field induced by diffraction, dispersion, or a nonlinear process that is proportional to pressure, are scale-invariant. 
In practice, an optical process in a gas medium, defined by a nonlinear effect and certain input parameters (pulse energy, gas pressure, focusing geometry), can be up- or down-scaled to different pulse energies without changing its general characteristics. Furthermore, our scaling formalism preserves the carrier-envelope phase (CEP), which only changes because of linear and nonlinear (e.g. self phase modulation) propagation effects, both of which are scale invariant. This implies that strongly CEP-dependent processes such as single attosecond pulse generation can be invariantly scaled. 
The scaling principle is illustrated in Fig.~\ref{fig:illustr} using the example of beam reshaping under the influence of nonlinear propagation and applied below to filamentation and attosecond pulse generation.

{\renewcommand{\arraystretch}{1.33}
\begin{table}
\centering
\begin{tabular*}{.48\textwidth}{L{3.0cm}C{2.8cm}C{1.6cm}}
\hline
\hline
 & Parameter & Scaled parameter\\
 \hline
Input parameter & & \\
\multirow{2}{*}{\quad Dimensions}&$z$ & $\eta^2 z$ \\
&$r$ & $\eta r$\\
\multirow{2}{*}{\quad Other parameters}&$\rho$& $\rho/\eta^2$\\
&$\varepsilon_{\textrm{in}}$& $\eta^2 \varepsilon_{\textrm{in}}$\\
\hline
Output parameters &  & \\
\quad General &$\varepsilon_{\textrm{out}}$& $\eta^2 \varepsilon_{\textrm{out}}$\\
\multirow{2}{*}{\quad Filamentation}&$p_{\mathrm{cr}}$&$\eta^2 p_{\mathrm{cr}}$\\
&$z_{\mathrm{cr}}$&$\eta^2 z_{\mathrm{cr}}$\\
\multirow{2}{*}{\quad HHG}&$\varepsilon_q$&$\eta^2 \varepsilon_q$\\
&$\Gamma_q$&$\Gamma_q$\\
\hline
\hline
\end{tabular*}
\caption{\textbf{Scaling relations:} This table lists the scaling relations derived in this work. $p_{\mathrm{cr}}$ and $z_{\mathrm{cr}}$ denote the critical power and the distance, at which an initially collimated beam collapses due to self-focusing. $\varepsilon_q$ and $\Gamma_q$ denote the harmonic pulse energy and the conversion efficiency into harmonic order $q$.}
\label{tab:sum}
\end{table} 

\subsection*{Filamentation}
A prominent example where several nonlinear propagation effects play a critical role, is filamentation \cite{CouaironPR2007}. A characteristic parameter influencing the onset of filamentation dynamics is the so-called critical power for self-focusing $p_{\mathrm{cr}} = N_{\mathrm{cr}}\lambda^2/ \,4\pi n_0 n_2$.
Here, $N_{\mathrm{cr}}$ is a constant depending on the spatial beam shape ($N_{\mathrm{cr}} = 1.896$ for a Gaussian transverse profile \cite{FibichOL2000}), $\lambda$ is the laser wavelength, $n_0$ the refractive index at the central frequency and $n_2$ the nonlinear refractive index. Since $n_2$ is to very good approximation, proportional to the gas density, $p_{\mathrm{cr}} \rightarrow \eta^2 p_{\mathrm{cr}} $, thus following the same scaling relation as $\varepsilon_{\textrm{in}}$, i.e. the critical power increases linearly with laser pulse energy. It can also be shown that the distance $z_{\textrm{cr}}$ at which an initially collimated laser beam collapses due to self-focusing, scales quadratically with the initial beam size (i.e. $z_{\textrm{cr}}   \rightarrow \eta^2 z_{\textrm{cr}}$) \cite{Marburger75}, confirming that the scaling transformations remain valid under nonlinear propagation conditions.

\begin{figure*}[tb!]
	\begin{center}
		\includegraphics[width=0.98\textwidth]{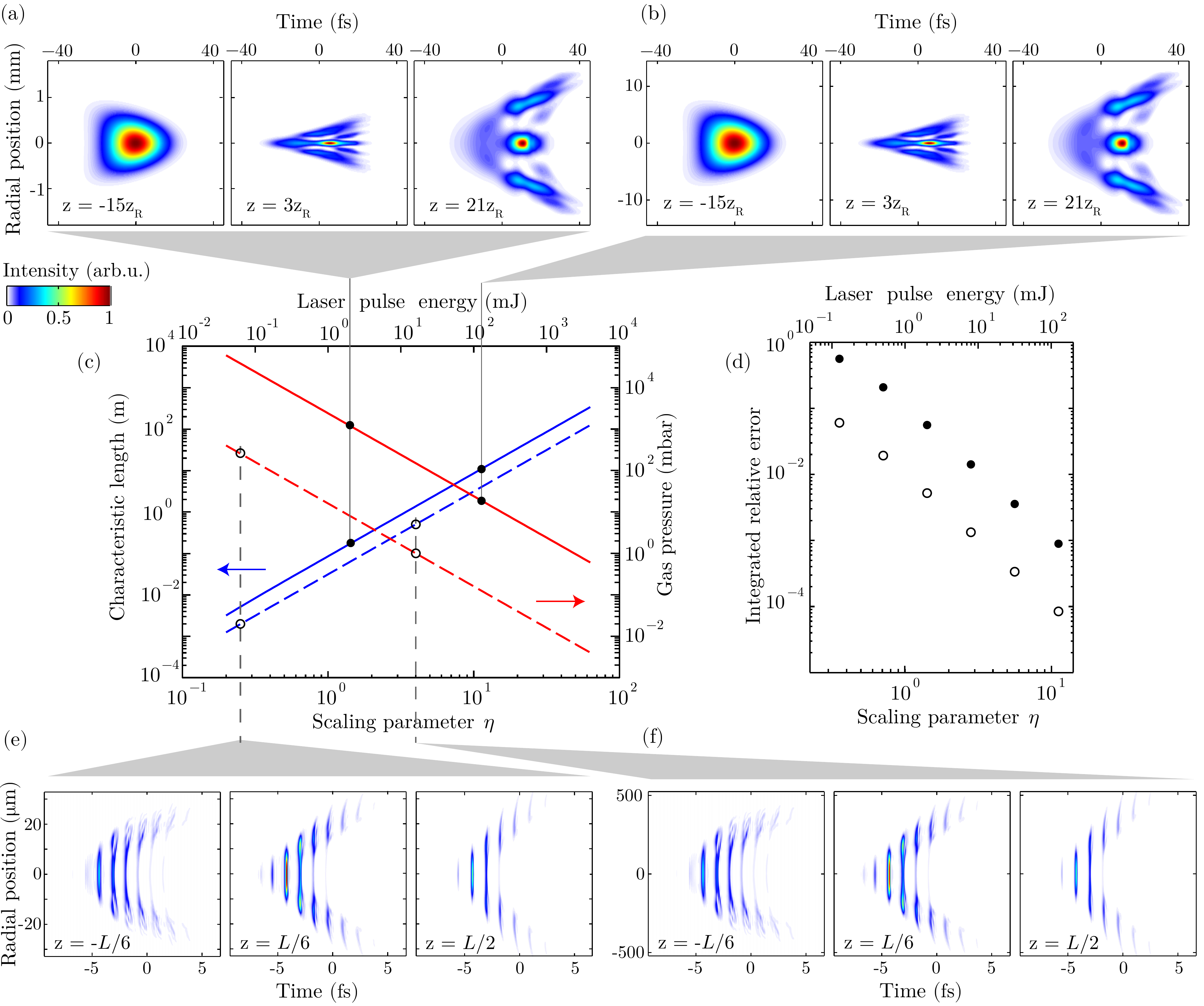}
		\setlength{\unitlength}{1mm}
	\end{center}
	\caption{\textbf{Scaling filamentation and HHG:} \textbf{a, b}, Simulated spatiotemporal intensity distributions (normalized individually) in a focused laser beam in Ar for three different positions along the propagation axis and two input parameter sets, scaled according to the presented scaling relations (\textbf{a}: $\tau = 20$\,fs, $\varepsilon_{\textrm{in}} = 2$\,mJ, $p = 1.2$\,bar, $W_0 = 40\,\mu$m, \textbf{b}: $\tau = 20$\,fs, $\varepsilon_{\textrm{in}} = 128$\,mJ, $p = 18.75$\,mbar, $W_0 = 320\,\mu$m).
\textbf{e, f}, Simulated spatiotemporal intensity distributions for high-harmonic emission (above 31.5\,eV) in Ar at three positions within the nonlinear medium (\textbf{e}: $\tau = 10$\,fs, $\varepsilon_{\textrm{in}} = 62.5\,\mu$J, $p = 256$\,mbar, $W_0 = 10.6\,\mu$m, $L = 2$\,mm, \textbf{f}: $\tau = 10$\,fs, $\varepsilon_{\textrm{in}} = 16$\,mJ, $p = 1$\,mbar, $W_0 = 169.6\,\mu$m, $L = 0.51$\,m). For both filamentation and HHG, the longitudinal position is specified with respect to the position of the geometrical focus; in \textbf{a} and \textbf{b} in units of the respective Rayleigh lengths and in \textbf{e} and \textbf{f} in units of the length of the generation medium $L$. 
\textbf{c}, characteristic length, i.e. filament and gas cell length, respectively (blue, left axis) and gas pressure (red, right axis) as a function of $\eta$ and $\varepsilon_{\textrm in}$. $\eta$ was arbitrarily set to unity for $\varepsilon_{\textrm{in}} = 1$\,mJ. \textbf{d}, Integrated relative scaling error for the filament scaling presented in \textbf{a} and \textbf{b} for intensity (dots) and fluence (circles) as defined in Methods.}
	\label{fig:filament}
\end{figure*}

We performed a more rigorous verification of our scaling model by numerically simulating filamentation with a state-of-the-art pulse propagation code (see Methods for details). 
Figure~\ref{fig:filament}\,(a) and (b) illustrate filamentation in Ar, using a 20\,fs input pulse centered at 800\,nm and two different parameter sets, where parameter set (b) corresponds to the up-scaled parameters ($\eta = 8$) of parameter set (a). In Fig.~\ref{fig:filament}\,(a) and (b), the spatiotemporal intensity distribution is shown for three positions along the optical axis. 
In both cases, typical filamentation characteristics like conical emission and temporal self-compression \cite{CouaironPR2007} can be observed.
Despite the very different pulse energies [$\eta^2 = 64$ times larger for parameter set (b)] and transverse scales ($\eta = 8$ times larger), only minor differences are visible demonstrating the validity of the scaling model for filamentation.
Figure~\ref{fig:filament}\,(c) illustrates how experimental parameters like input energy, gas pressure and filament length (defined here as the propagation length over which the intensity on the optical axis exceeds $5\cdot 10^{13}$\,W/cm$^2$) scale with $\eta$.

Figure~\ref{fig:filament}\,(d) shows a numerically extracted relative scaling error, representing the deviation from perfect scalability for output intensity (dots) and fluence (circles) as a function of $\varepsilon_{\textrm{in}}$. For each pulse energy, the error was calculated by comparing the output intensity (or fluence) to that obtained with four times larger pulse energy (see Methods for details).
While the scaling error is negligibly small for pulse energies well above 1~mJ, thus indicating no fundamental upper scaling limit, a clear deviation from perfect scaling appears for small pulse energies. These deviations can be mainly attributed to avalanche ionization (see SI).
  
\subsection*{Attosecond Pulse Generation}

As a second example, we consider HHG in gases, which leads to the formation of attosecond pulse trains or single attosecond pulses. 
Although both cases are easily encompassed in our scaling model, for simplicity we concentrate on pulse trains.
HHG in an extended nonlinear medium can be described in two steps: first, the laser pulse propagates through the nonlinear medium, inducing a polarization $\hat{P}_q=2d_{q} \rho$, at multiple, odd-order harmonic frequencies, where $d_{q}$ is the single atom nonlinear dipole moment.
Second, the harmonic field $\hat{E}_q$ is generated from the induced polarization. The propagation of $\hat{\cal E}_q= \hat{E}_q \exp(-i \omega z/c)$, where $\omega$ now denotes the harmonic frequency, can be described by equation~\eqref{eq:fme1}, with  $\hat{\cal P}_{\text{NL}}$ being replaced by $\hat{\cal P}_q$.
Since both the fundamental and the harmonic fields follow scale-invariant propagation equations, HHG is invariant under the scaling transformations.
Consequently the harmonic output pulse energy $\varepsilon_q$ follows the same scaling $\varepsilon_{q}  \longrightarrow  \eta^2 \varepsilon_{q}$.
This implies that the conversion efficiency $\Gamma_{q} =\varepsilon_q/\varepsilon_{\textrm{in}}$ 
is scale-invariant. In other words, the same conversion efficiency can be expected for HHG driven by intense laser pulses, with a loose focusing geometry as well as by much weaker laser pulses, with tight focusing geometry, as recently discussed in Refs.~\cite{HeylJPB2012,RothhardtNJP2014}. 

We verified the scalability by simulating HHG in Ar, using a simulation code that includes both laser and XUV field propagation effects, see Methods for details. 
The dipole response was calculated using the Strong Field Approximation \cite{LewensteinPRA1994}.
Figure~\ref{fig:filament}\,(e) and (f) illustrate HHG using 10\,fs laser pulses centered at 800\,nm.
Similar as in Figure~\ref{fig:filament}\,(a) and (b), spatiotemporal intensity maps are displayed, showing the evolution of the total field build-up along the nonlinear medium for two parameter sets, differing by $\eta = 16$ (a factor 256 in input energy!). The total field above 31.5 eV (i.e. from the 21st harmonic) is represented. It exhibits a train of ultrashort, attosecond pulses. The generation parameters led to significant pulse reshaping effects due to plasma formation, implying that the generation conditions were not optimized for efficient HHG. The high intensity leads to divergent, ring-like emission except at the rising edge of the laser pulse. Again, an almost perfect scaling behavior can be observed, confirming $\varepsilon_q \longrightarrow 256\,\varepsilon_q$.

\subsection*{Experimental verification}

To verify the scaling experimentally, we performed pulse compression experiments via filamentation in gases with 20\,fs input pulses (FWHM) centered at 800\,nm.
The pulse energy was varied in the range of $\varepsilon_i=0.12-2.7$\,mJ and spherical mirrors with focal lengths $f=0.5-2.5$\,m were used to focus into an argon-filled tube with a length approximately twice the respective focal length. We span a pulse energy range of $\sim$25, the highest energy being limited by laboratory space constraints.
The pulses emerging from the filament were compressed with chirped mirrors and fused silica wedges and characterized using the dispersion-scan technique \cite{MirandaOE2012} (see Methods for details).
Figures~\ref{fig:experiment}(a) and (b) show temporal intensity as well as spectral amplitude and phase for six different input pulse energies. 
For the shortest focal lengths (lowest pulse energy) gas pressure and pulse energy were optimized for maximum spectral broadening and good compressibility, while avoiding multiple filamentation.
For all other measurement points, focal length and gas pressure were adjusted according to the scaling relations, while the pulse energy was used as a free parameter to optimize the output spectrum, resulting in input pulse energies very close to the scaling prediction.
All employed experimental parameters together with fits visualizing the expected scaling trend are displayed in Figure ~\ref{fig:experiment}(c).
The post-compressed pulse duration as well as the overall characteristics are very similar for all six cases, indicating very good scalability of all relevant linear and nonlinear propagation processes within the employed parameter range.   

\begin{figure*}[htb!]
	\begin{center}
		\includegraphics[width=0.96\textwidth]{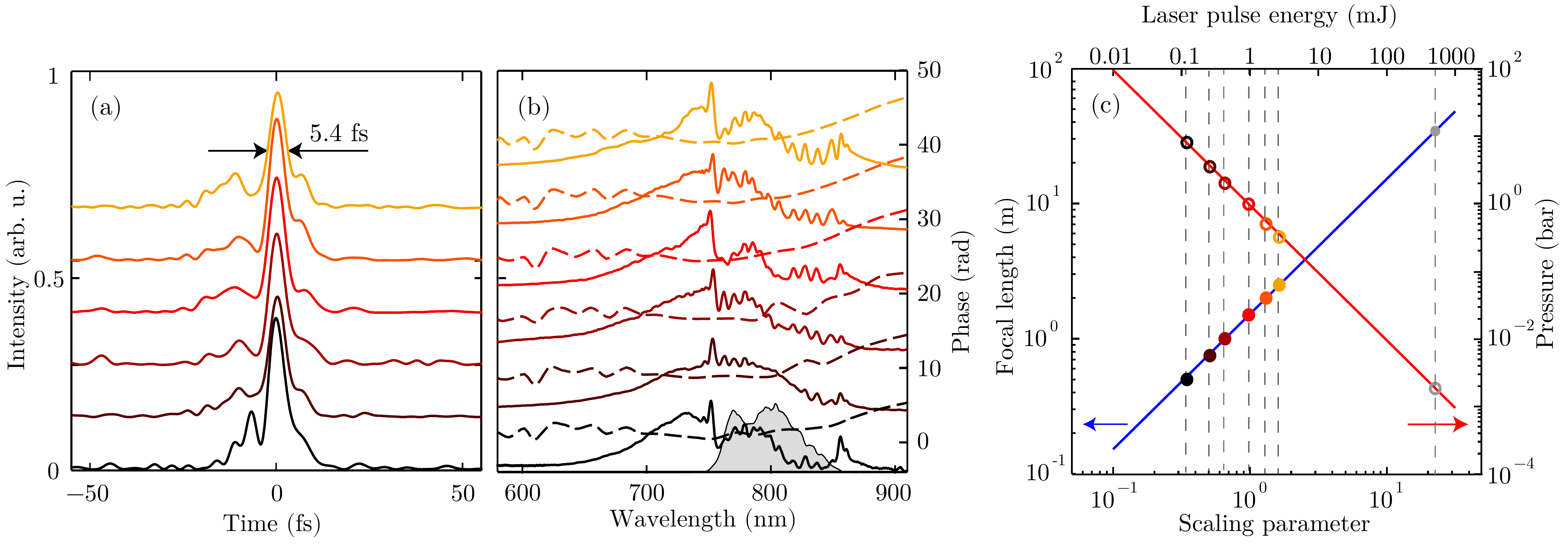}
		\setlength{\unitlength}{1mm}
	\end{center}
	\caption{\textbf{Experimental filament scaling:} \textbf{a,b}, Measured temporal intensity profiles as well as spectral amplitude (\textbf{b}, solid lines) and phase (\textbf{b}, dashed lines) for six different parameter sets, shown in \textbf{c}. For better visualization, the plotted datasets are vertically off-set from each other. The measurement was performed selecting the broadband radiation on the optical axis more than a focal length distance behind the filament. For reference, the input spectrum (gray shaded area) is shown in \textbf{b}. The solid lines in \textbf{c} represent fits to the experimental data points, as defined by the presented scaling relations, indicating the expected scaling performance for input laser pulse energies within and beyond the measured parameter range. The gray data points in \textbf{c} visualize the extrapolated parameters shown in Table~\ref{tab:extrapolate}.}
	\label{fig:experiment}
\end{figure*} 

Up (and even down-) scaling HHG has been investigated previously, albeit in a phenomenological way, and in many cases without changing consistently all relevant parameters included in our scaling formalism. 
To make use of high input energies, loose focusing geometries have been implemented since the early days. Although it was often realized that in these conditions the use of long media (and low pressures) led to higher XUV pulse energies, to our knowledge, no rigorous understanding for this experimental observation has been put forward. 
In the other direction, tight focusing geometries, necessary for HHG with laser systems with low input energy (a few tens of $\mu$J), have often been implemented and found to be detrimental for phase matching, and thus for the conversion efficiency. A recent experiment using a short medium at high pressure \cite{RothhardtNJP2014}, however, shows a similar conversion efficiency as with loose focusing, in perfect agreement with our scaling predictions.   

\subsection*{Discussion and generality of the scaling principle}
{\renewcommand{\arraystretch}{1.33}
\begin{table}[b]
\centering
\begin{tabular*}{.48\textwidth}{L{2.3cm}C{1.1cm}C{1.1cm}C{1.1cm}C{1.1cm}}
\hline
\hline
& $\varepsilon_{\mathrm{in}}$ & $\varepsilon_{\mathrm{out}},\varepsilon_{q}$ & $p$ & $f$\\
\hline
Filamentation & (mJ) & (mJ) & (mbar)& (m)\\
\quad	Typical & 1 & 0.1 & 980 & 1.5 \\
\quad Up-scaled & 500 & 50 & 1.96 & 33.5 \\
\hline
HHG & (mJ) & (nJ) & (mbar)& (m)\\
\quad Typical & 1.5 & 15 & 15 & 1\\
\quad Up-scaled & 500 & 5000 & 0.045 & 18.3 \\
\quad Down-scaled & 0.01 & 0.1 & 2300 & 0.08\\
\hline
\hline
\end{tabular*}
\caption{\textbf{Extrapolation of typical parameters for filamentation and HHG}: The experimentally well explored regime around $\varepsilon_{\mathrm{in}} \approx 1$\,mJ (denoted as typical) is up- and (for HHG) down-scaled. An input pulse length of 20\,fs is taken into account for filamentation and 40\,fs for HHG, with Ar as nonlinear medium. The beam diameter (FWHM) before focusing is in both cases $\sim 7$\,mm.}
\label{tab:extrapolate}
\end{table} 

We illustrate the scaling possibilities and experimental challenges for the two phenomena discussed in Table~\ref{tab:extrapolate}. Starting from typical experimental parameters corresponding to $\varepsilon_{\mathrm{in}} \sim 1$\,mJ, we apply our scaling relations both for filamentation and HHG up to $\varepsilon_{\mathrm{in}} = 500$\,mJ and, for HHG, down to $\varepsilon_{\mathrm{in}} = 10\,\mu$J and calculate the expected values for output pulse energy, gas pressure and focal length. In the case of filamentation, we use the parameters of the experiment presented above and assume that $\varepsilon_{\mathrm{out}} = 0.1 \varepsilon_{\mathrm{in}}$ for the central compressed part of the filament. In the case of HHG, we start from values close to those reported in \cite{ConstantPRL1999} with a conversion efficiency in Ar equal to $10^{-5}$ (see also \cite{RudawskiRSI2013}). 
These examples illustrate the feasibility of both up and down-scaling.  
Even though long geometries need to be implemented, few-cycle laser pulses and attosecond XUV pulses with unprecedented energies are within reach. Conversely, high gas densities and very tight focusing geometries are required for the efficient generation of attosecond pulses at MHz repetition rates \cite{RothhardtNJP2014}.

Our scaling model does not indicate any limitation for up-scaling. However, for down-scaling, several effects leading to deviations from perfect scalability can be identified. 
First, non-paraxial propagation effects arise at very tight focusing geometries (typically at numerical apertures $\gtrsim$ 0.3). Second, the not perfectly linear dependence of $K(\omega,\rho)$ and possibly $\hat{\cal{P}}_{\textrm{NL}}$ on the gas density [see equation~\eqref{eq:fme1}] as well as the weak dependence of $1/k(\omega,\rho)$ on the density contribute to increasing deviations from perfect scaling (see SI for details). 
Finally, at high ionization levels and high densities, avalanche ionization, a process that critically depends on plasma dynamics and that is not scalable according to our model, can set strict limitations (see SI).
In extreme conditions, the generated plasma can become opaque (for $p \gtrsim 70$\,bar at 800\,nm and room temperature, assuming a totally singly-ionized medium) and, for processes like HHG, the single atomic response can be affected by the presence of neighboring atoms \cite{StrelkovPRA2005}. However, we estimate that these effects do not play a major role within the parameter ranges typically employed, for example, for HHG in gases and for filamentation [see also Fig.~\ref{fig:filament}(d)].

The presented scaling framework is very general and applies to other processes, involving linear or nonlinear electromagnetic wave propagation in gases. The key condition determining if a nonlinear process is scale-invariant is the proportionality $\hat{\cal{P}}_{\textrm{NL}} \propto \rho$. Nonlinear processes which critically depend on plasma dynamics such as avalanche ionization or the acceleration of electrons in relativistic light
fields \cite{EsareyRMP2009} are thus not fully scalable according to our derived formalism. Furthermore, for processes that make use of the plasma as source of secondary emission, the frequency dependence of the secondary radiation upon gas density induces a non-negligible departure from $\hat{\cal{P}}_{\mathrm{NL}} \propto \rho$ and thus from scale-invariance.
Nonlinear interactions that are scalable to a very good approximation include self-focusing, self-phase modulation, wave mixing, as well as field ionization, plasma defocusing and processes involving stimulated Raman scattering.  
Similar scaling principles can also be applied for pulse propagation in waveguides such as hollow capillaries \cite{BohleLPL2014}. 
We expect our results to be of great interest for ultrafast science since we show how to extend different nonlinear methods to the new parameter regimes provided by today's state-of-the-art femtosecond laser technology. Our findings are currently being applied to the design of an up-scaled, next-generation attosecond source, for the European facility ELI-ALPS (Extreme Light Infrastructure - Attosecond Light Pulse Source).

\hyphenation{Post-Script Sprin-ger}

\section*{Methods}
\subsection*{Numerical Methods}
\noindent{\em Filamentation:}
For the filamentation simulations, a uni-directional nonlinear envelope equation was numerically integrated \cite{ArnoldNJP2010}. This type of equation is valid for pulses down to the single-cycle regime \cite{ BrabecPRL1997}. Note that the use of an envelope equation is not in contradiction to our discussion; the scaling can be formally shown both for envelope and field equations. The propagation equation follows from the nonlinear Helmholtz equation, adapting the method suggested by Feit and Fleck to introduce uni-directionality \cite{FeitAPL1974}. Argon is used as nonlinear medium and dispersion is treated by a Sellmeier-type expression for the linear refractive index \cite{Dalgarno1966}. Self-focusing originating from a Kerr-type nonlinearity is described by a nonlinear refractive index $n_2 = 0.98 \times 10^{-19}$~cm$^2$W$^{-1}$ for Ar at atmospheric pressure \cite{TempeaOL1998}.
The interaction with the generated plasma, i.e. absorption and plasma defocussing, is treated in terms of a current density, whereas free electrons are generated via field ionization, numerically described within the generalized Keldysh-PPT theory (Perelomov-Popov-Terent'ev) \cite{ KeldyshJETP1965, PerelomovJETP1966}. Avalanche ionization is included via a Drude-type model with a phenomenological collision time of $\tau_C = 190$~fs at atmospheric pressure \cite{CouaironJModOpt2006}.

\noindent{\em HHG:}
The numerical HHG simulations are performed in two steps: in the first step the generating laser field is propagated through the cylindrically symmetric gas medium, taking into account the time- and space-dependent neutral and plasma dispersion and optical Kerr effect, as described in Ref.~\cite{TosaPRA2005}. Then, in the second step the harmonic field is propagated using the generated high-order polarization as source term calculated using the strong-field approximation \cite{LewensteinPRA1994}. For the propagation of the harmonic field neutral dispersion and absorption are taken into account. 
The wave equations for both fundamental and harmonic fields are derived in paraxial approximation. In both equations a nonlinear source term proportional to the gas density is assumed. 
For the fundamental field absorption resulting from optical field ionization is taken into account (also scaling linearly with gas density), while neutral and plasma absorption are neglected.
The numerical code has been verified by comparison with experimental results obtained with the intense harmonic beam line at Lund University.

\subsection*{Experimental methods}
The experiments were performed with a Titanium:Sapphire chirped pulse amplification laser chain, delivering up to 5\,mJ pulses, with 20\,fs duration (FWHM), centered around 800\,nm, at a repetition rate of 1\,kHz. The beam size was adapted to a diameter ($1/e^2$ of the intensity profile) of approx. 11\,mm with a reflective telescope, followed by a 10\,mm aperture. The filaments were generated inside an Ar-filled tube using spherical mirrors of different focal lengths (\(f = 0.5, 0.75, 1, 1.5, 2, 2.5\)\,m). The tube was terminated with thin (0.5\,mm) fused silica windows in order to minimize the impact of dispersion and undesired nonlinear interaction. The Ar pressure was set according to the scaling law ($p = 8, 3.5, 2, 0.98, 0.50, 0.32$\,bar), while the pulse energy was fine-tuned with a $\lambda/2$-plate followed by a transmissive polarizer in order to match the super-continuum spectra originating from the filament to best possible agreement for the different combinations of pulse energy, Ar pressure, and focal length. For the lowest input pulse energy, a second polarizer plate was installed in order to improve the polarization contrast.
Behind the filament tube, the bright center of the filament was selected with an aperture and send to a pulse compressor, comprised of movable fused silica wedges and chirped mirrors in double-angle configuration (Ultrafast Innovations). The pulses were characterized using the dispersion scan technique (d-scan) \cite{MirandaOE2012}. The d-scan is based on recording the second harmonic spectrum as function of dispersion, in our case realized by moving one of the wedges. The spectral phase is retrieved with a numerical algorithm that iteratively reproduces the measured d-scan trace (second harmonic spectrum vs. wedge insertion) based on a measured fundamental spectrum, see Figure~1 in the SI. For the temporal pulse retrievals shown in Figure~\ref{fig:experiment}(a), a wedge insertion was chosen corresponding to a minimized pulse duration and best agreement with pulse profiles retrieved for other parameter combinations. Note that an offset third-order phase remains uncompensated in the employed compression scheme, as visible in the d-scan traces, showing a clear variation of spectral frequency distribution as a function of wedge insertion.

\subsection*{Scaling error calculation}

The normalized relative scaling errors for intensity $I_n = I_n(r,t)$ and fluence $F_n = F_n(r,z) = \int_{-\infty}^{\infty} I_n(r,t,z)\textrm{d}t$ displayed in Figure~\ref{fig:filament} were calculated for the input parameter set ($\varepsilon_{in}^n$, $p_n$, $f_n$) and referenced to the corresponding parameter set with four times higher input pulse energy, all other parameters being adjusted according to the scaling relations, i.e. ($\varepsilon_{in}^{n-1} = 4 \varepsilon_{in}^n$, $p_{n-1} = p_n/4$, $f_{n-1} = 2f_n$). For the intensity and the fluence the normalized errors can be written as:
\begin{equation*}
\nu_{\mathrm{int}}^n = \frac{2 \int_{\infty} dt \int_0^{\infty} \textrm{d}r\,r \left|I_{n} - I_{n-1}\right|}{\int_{\infty} \textrm{d}t \int_0^{\infty} \textrm{d}r\,r I_{n} +  \int_{\infty} \textrm{d}t \int_0^{\infty} \textrm{d}r\,r I_{n-1}},
\label{eq:errorint}
\end{equation*} 
\begin{equation*}
\nu_{\mathrm{flu}}^n = \frac{2 \int_{\infty} \textrm{d}z \int_0^{\infty} \textrm{d}r\,r \left|F_{n} - F_{n-1}\right|}{\int_{\infty} \textrm{d}z \int_0^{\infty} \textrm{d}r\,r F_{n} +  \int_{\infty} \textrm{d}z \int_0^{\infty} \textrm{d}r\,r F_{n-1}}.
\label{eq:errorflu}
\end{equation*}
The integrals where computed numerically on a finite grid in $r$ and $z$ using a normalized grid size such that $\Delta r/W_0 = \mathrm{constant}$ and $\Delta z/z_{\mathrm{R}} = \mathrm{constant}$ with $\Delta r$ and $\Delta z$ denoting the grid unit size.    

\section*{Acknowledgments}
We thank P.~Rudawski, B.~Manschwetus, S.~Maclot and P.~Johnsson for the experimental verification of the numerical HHG code and their contribution to discussing the scaling of HHG.

This work was supported by the European Research Council (PALP), the Marie Curie ITN MEDEA, the Knut and Alice Wallenberg Foundation, the Swedish Research Council, the European Union and the European Regional Development Fund (GOP-1.1.1-12/B-2012-0001), the Hungarian Scientific Research Fund (OTKA project NN107235), E02/2014 (PULSE-PROPAG) and PN-II-ID-PCE-2012-4-0342.

\section*{Author contributions}

C.~M.~H. conceived the idea and wrote major parts of the manuscript, C.~L.~A. performed the filamentation simulations, K.~K., V.~T., E.~B. and K.~V., performed the HHG simulations, C.~M.~H., H.~C.~A., M.~M., M.~L. and C.~L.~A. performed the experiments. C.~M.~H, C.~A., A.~C., and A.~L. and set up the analytical scaling formalism. All authors contributed to discussing the results and the writing of the article.  

\end{document}